\documentclass[nofootinbib,aps,prd,superscriptaddress,showpacs, preprint]{revtex4-1}
\usepackage{graphicx}
\usepackage{epstopdf}
\usepackage{amsmath}
\usepackage{amsfonts}
\usepackage{amssymb}
\usepackage{latexsym}
\usepackage{hyperref}
\usepackage{xcolor}
\usepackage[colorinlistoftodos]{todonotes}

\begin{document}

\title{Decay photons from the ALP burst of type-II supernovae }

\author{J. Jaeckel}\email{jjaeckel@thphys.uni-heidelberg.de}
\affiliation{Institut f\"ur theoretische Physik, University of Heidelberg, Heidelberg, Germany}

\author{P.C. Malta}\email{malta@thphys.uni-heidelberg.de}
\affiliation{Institut f\"ur theoretische Physik, University of Heidelberg, Heidelberg, Germany}
\affiliation{Centro Brasileiro de Pesquisas F\'{i}sicas (CBPF), Rio de Janeiro, Brazil}

\author{J. Redondo}\email{jredondo@unizar.es}
\affiliation{University of Zaragoza, Zaragoza, Spain}
\affiliation{Max Planck Institut f\"ur Physik, Munich, Germany}


\begin{abstract}

We determine limits from SN 1987A on massive axion-like particles with masses in the 10 keV - 100 MeV range and purely coupled to two photons. 
Axion-like particles produced in the core collapse escape from the star and decay into photons that can be observed as a delayed and diffuse burst. We discuss the time and angular distribution of such a signal. Looking into the future we also estimate the possible improvements caused by better gamma-ray detectors or if the red supergiant Betelgeuse explodes in a supernova event.  

\end{abstract}


\maketitle


\section{Introduction}

\indent

Many beyond the Standard Model scenarios include new massive (pseudo-)scalars -- dubbed {\it axion-like particles} (ALPs) -- among their particle spectrum (see e.g. Refs.~\cite{Coriano,Dias,Dubovsky} and Refs.~\cite{Cicoli, Ringwald_0, JJ} for recent reviews). The name originates from their similarity to the axion of the Peccei-Quinn solution to the strong CP problem\cite{PQ,weinberg,wilczek}. Contrary to the usual QCD axion, that also couples to gluons, ALPs may solely interact with two photons,
\begin{equation}
\label{coupling}
{\mathcal{L}}_{\rm int}\supset\frac{g_{a\gamma\gamma}}{4}aF_{\mu\nu}\tilde{F}^{\mu\nu},
\end{equation}
where $a$ denotes the ALP and $g_{a\gamma\gamma}$ is its coupling constant with dimension of inverse energy, which is often linked to an underlying scale of new physics $f_{a}$ via $g_{a\gamma\gamma}\sim\frac{\alpha}{2\pi}\frac{1}{f_{a}}$. In this paper we will focus on a pure coupling to photons\footnote{The main reason for this is simplicity, but also that this is an often considered test model. Nevertheless, other interactions -- in particular new decay modes -- could be included by an appropriate recasting using an adapted decay length and branching ratio into two photons along similar lines as discussed in Ref.~\cite{duffy}.} as given by Eq.~\eqref{coupling}. In contrast to the QCD axion, in the more general case of ALPs there is no fixed relation between mass and coupling: these are taken as completely independent parameters.

The coupling to photons allows the production of ALPs in stellar cores via the Primakoff mechanism. Even if this coupling is extremely small, a sizeable amount of ALPs can be produced in the stellar bulk and have observable consequences. 
Here our aim is to study very weakly coupled ALPs produced in type-II supernovae (SN) and the effects of their decay into photons outside of the star. The objective is to determine which regions in the $g_{a\gamma\gamma} - m_a$ space are allowed (or not) by the exquisite sensitivity that gamma-ray satellites can have to measure this photon burst. As we will detail in Section~\ref{basics}, this process is most effective in the 10 keV - 100 MeV mass range, where masses are smaller than the temperature of the SN core but large enough as to make the decay rate outside the star sizeable.

\begin{figure}[t!]
\centering
\includegraphics[angle=0,width=0.7\textwidth]{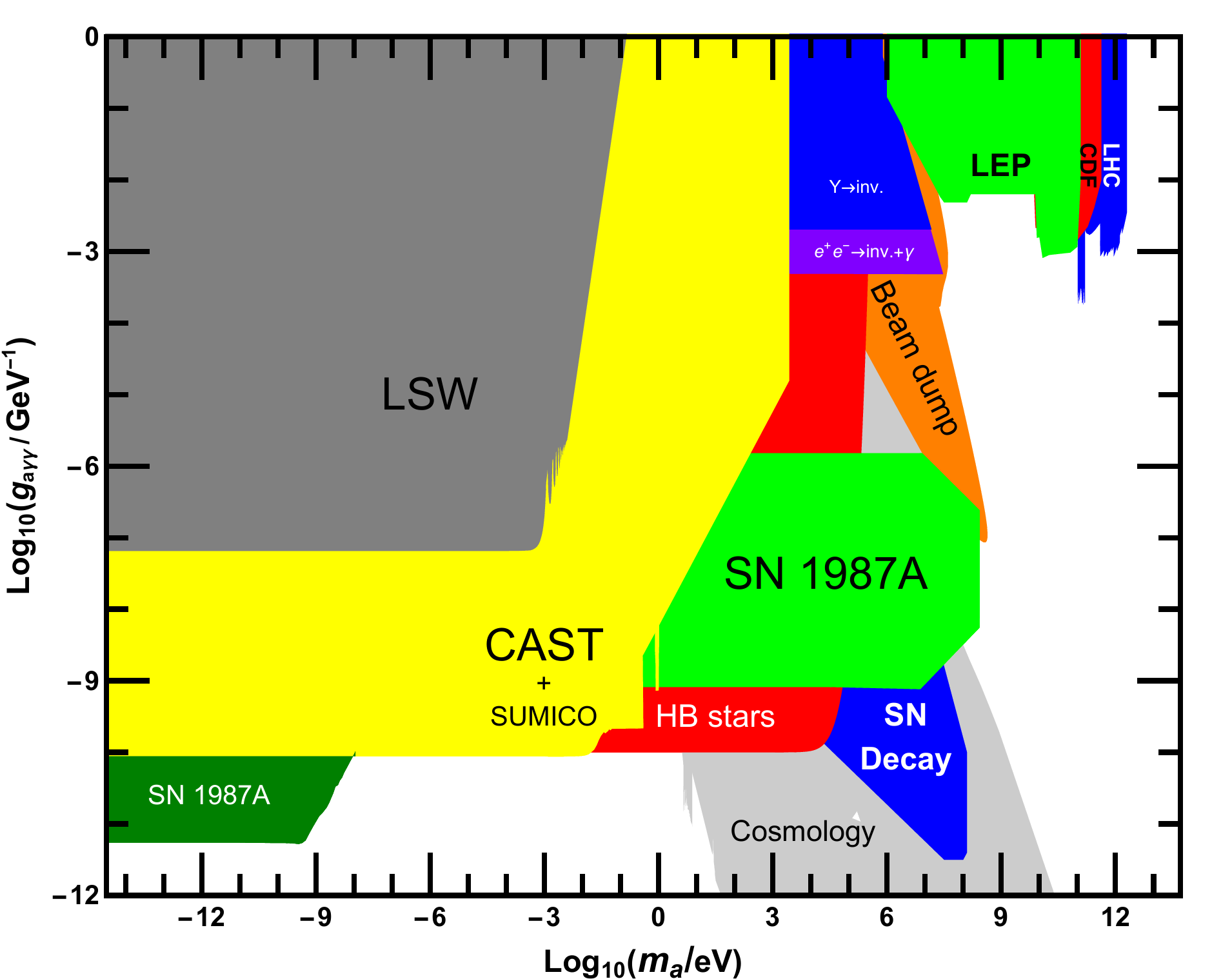}
\caption{Excluded regions in ALP parameter space (figure adapted from Refs.~\cite{Redondo:2008en,JJ,Alekhin:2015byh,Jaeckel:2015jla} with added limits from Refs.~\cite{cadamuro,Hewett:2012ns,Jaeckel:2012yz,Mimasu:2014nea,ringwald,Millea,Dobrich:2015jyk}). Our bound is shown in dark blue (``SN decay").} \label{exc_plot}
\end{figure}

Our main result is the excluded region labelled ``SN decay" in the ALP parameter space of Fig.~\ref{exc_plot}. In this region our arguments provide better limits than existing laboratory and astrophysical constraints. ALPs in this parameter range can have a strong impact in cosmology if their relic density is as large as the thermal one, but this depends on the maximum temperature of the big bang~\cite{cadamuro} which is, at the moment, unknown (cf. Section~\ref{conclusion}). 

We focus on SN 1987A, which has already been exploited to derive a variety of limits on ALPs. Perhaps the simplest one arises from the energy loss implied by significant ALP emission, which would reduce the measured neutrino burst below the $\sim 10$ s observed by neutrino detectors~\cite{Raffelt,Raffelt1} (light green region labelled SN 1987A in Fig.~\ref{exc_plot}). For very light ALPs with masses below $m_a < {\rm few} \times 10^{-10} \, {\rm eV}$ a better limit can be obtained by taking into account that ALPs emitted from the supernova can convert into photons in the magnetic field of the galaxy~\cite{Stodolsky, Sikvie}, but no gamma-ray signal was ever detected after SN 1987A~\cite{Brockway,Toldra,Angelis, Simet, Serpico,ringwald} (dark green region labelled SN 1987A)\footnote{For a future supernova the sensitivity could be improved employing Fermi-LAT~\cite{Osc}.}.  For heavier ALPs this does not work because the reconversion into photons is strongly suppressed.

For sufficiently heavy ALPs with masses in the 10 keV - 100 MeV region however, another process becomes possible: the decay into two photons. This possibility was analysed in Ref.~\cite{duffy} assuming that ALPs are produced in the SN core via a direct ALP coupling to nucleons. In this paper we consider a less model-dependent case in which the photon coupling,  Eq.~\eqref{coupling}, is responsible for both ALP emission and decay. Moreover, we improve on several aspects of the calculation\footnote{In Section~\ref{sec_desc_sim} we briefly discuss the discrepancies between our results and those from Ref.~\cite{duffy})}. It is very important to notice that not all photons can reach Earth. We improve the estimate of the number that do so by employing a numerical simulation, in particular carefully treating the time delay and the angular distributions.

In addition to SN 1987A we also consider possible future SNe events. For concreteness, and also because it may produce spectacular effects, we entertain the core collapse of the red supergiant Betelgeuse. This is particularly interesting since its distance to Earth is only $\sim 200$ pc ($\sim 650$ ly), therefore much closer than SN 1987A (at 51.4 kpc $\sim$ 170000 ly).

The paper is organised as follows. In Section~\ref{basics} we discuss the essentials of the production mechanism, the subsequent decay and the geometrical features relevant for our analysis. In Section~\ref{simulation} we briefly describe our numerical simulation and the obtained results for the detectable fraction of signal photons as well as their distribution in time and angle. In Section~\ref{limits} we use SN 1987A to obtain concrete limits and discuss the potential sensitivity if Betelgeuse goes supernova. We summarize and conclude in Section~\ref{conclusion}.


\section{Setting up the analysis} \label{basics}

\indent

ALPs are produced in the core of the SN via the Primakoff process, $\gamma + p \rightarrow p + a$, whereby a thermal photon is converted into a pseudo-scalar in the presence of the external electromagnetic field provided by the charged particles in the medium (plasma). The typical energies of the produced ALPs are of the order of the core temperature and are in the $\sim 100 \, {\rm MeV}$ range. Recently the associated ALP energy spectrum has been calculated with detailed account of the production process in core-collapse SNe in Ref.~\cite{ringwald} and we will use the ensuing ALP production rate to estimate the ALP-originated photon fluence, i.e., the number of gamma-ray photons per unit area, on Earth.

Despite the core being extremely dense ($\sim 10^{14} \, {\rm g/cm}^3$), due to the smallness of the coupling to two photons, the ALPs escape the SN essentially unimpeded (cf. Eq.~\eqref{r_star}) and are emitted isotropically. As already mentioned, this is an effective energy-loss mechanism that in itself constrains the parameter space of ALPs~\cite{Raffelt,Raffelt1}. However, the couplings we are interested in are so weak that the energy taken by the emitted ALPs does not affect the core collapse. Rather, they contribute a tiny perturbation to the standard picture of the collapse. Nevertheless, our ALPs can decay outside the SN producing $\gtrsim$ 10 MeV photons, which is much larger than the typical photon energy in the outskirts of the SN. These gamma rays have no standard background, so we are sensitive to a much smaller flux.

The ALP decay rate is~\cite{PDG}
\begin{equation}
\label{LT}
\Gamma_{a\gamma\gamma} = \frac{g^2_{a\gamma\gamma}m^3_a}{64\pi},
\end{equation}
which results in a decay length for the ALPs given by
\begin{eqnarray}
\label{decaylength}
\ell_{\rm ALP}=\frac{\gamma v}{\Gamma_{a\gamma\gamma}}
\!&=&\!\frac{E_{a}}{m_{a}}\sqrt{1-\frac{m^{2}_{a}}{E^{2}_{a}}}\frac{64\pi}{g^{2}_{a\gamma\gamma}m^{3}_{a}}
\\\nonumber
\!&\approx&\! 4\times 10^{13}\,{\rm m}\left(\frac{E_{a}}{100\,{\rm MeV}}\right)
\left(\frac{10\,{\rm MeV}}{m_{a}}\right)^4\left(\frac{10^{-10}\,{\rm GeV}^{-1}}{g_{a\gamma\gamma}}\right)^{2}
\\\nonumber
\!&\approx&\!4\times 10^{-3}\,{\rm ly}\left(\frac{E_{a}}{100\,{\rm MeV}}\right)
\left(\frac{10\,{\rm MeV}}{m_{a}}\right)^4\left(\frac{10^{-10}\,{\rm GeV}^{-1}}{g_{a\gamma\gamma}}\right)^{2}
\end{eqnarray}
and, for the masses and couplings we are interested in, the decay length is large, but still allows for a sizeable number of ALPs to decay between SN 1987A and Earth ($d_{\rm SN} \sim 51 \, {\rm kpc}$).

Therefore, ALP decay is the relevant gamma-ray production mechanism to be considered and we may expect a flux of ALP-originated photons on Earth. Once we have the ALP production rate we may convolute it with the decay probability to obtain the fluence of ALP-originated photons at the detector. By obtaining (upper) limits on the gamma-ray fluence shortly after the observation of SN 1987A, we may constrain $g_{a\gamma\gamma}$ and $m_a$ by demanding that the number of ALP-originated photons arriving at the detector does not exceed what was observed~\cite{Chupp,Woosley}.

\begin{figure}[t!]
\centering
\includegraphics[angle=0,width=0.5\textwidth]{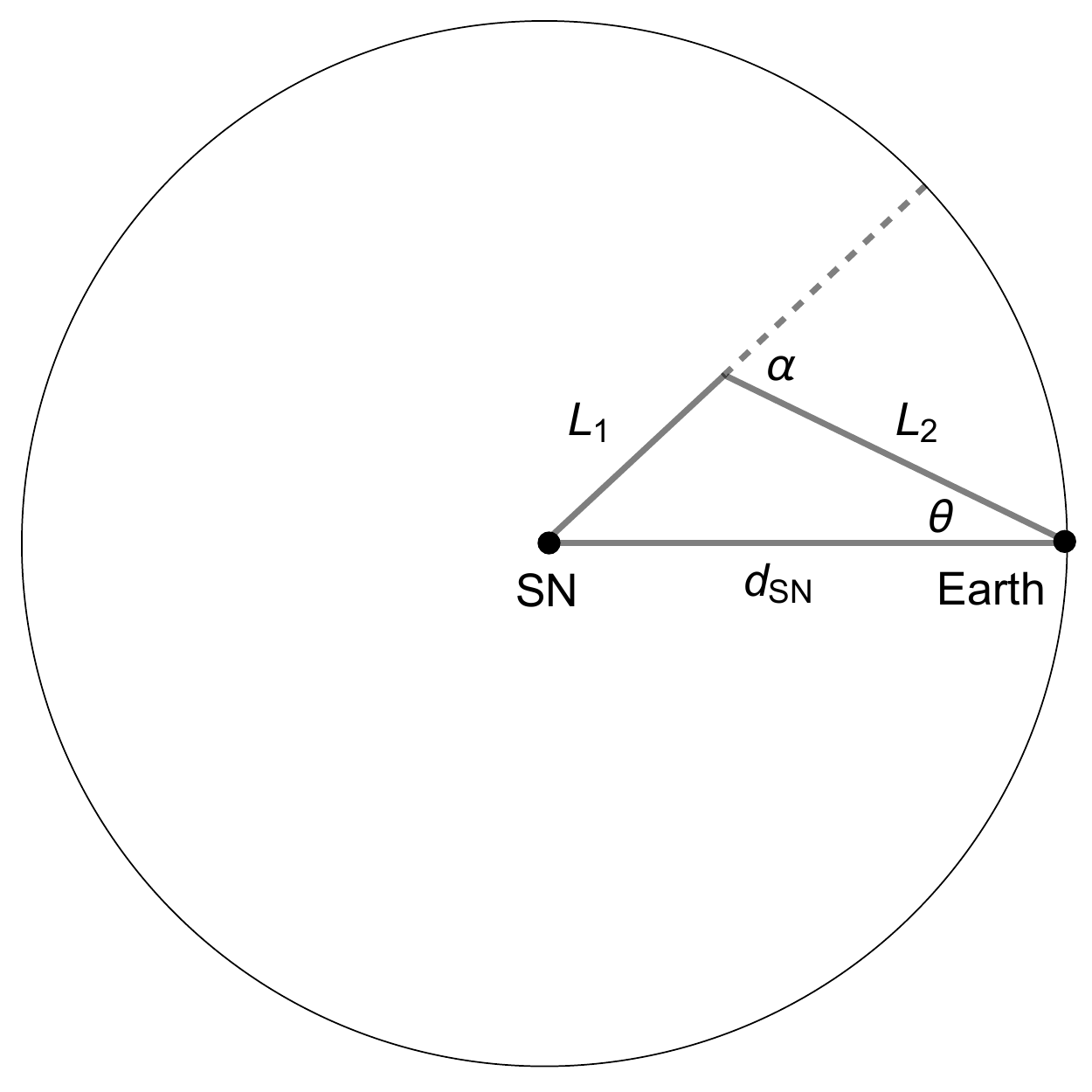}
\caption{Sketch of the geometry involved in the ALP production at the SN, its path (length $L_1$) until decay with an angle $\alpha$ and subsequent propagation of the photon (length $L_2$) until Earth, where it is detected under an angle $\theta$ (the angles and distances are exaggerated for the sake of clarity). Similar considerations are valid for the second photon.} \label{geometry}
\end{figure}

Another important point to be considered is that, having significant masses and being typically produced with energies $\sim {\rm few}\times10$ MeV, ALPs emitted from the SN core have appreciable -- but not enormous -- Lorentz boost factors ($\gamma = E_a/m_a$). This has important consequences. Firstly, the angle between the two photons and hence also between the original propagation direction $\alpha$ of the parent ALP is non-vanishing (see Eq.~\eqref{dist_ang}),
\begin{equation}
\sin\alpha \sim \gamma^{-1}, \label{dec_ang}
\end{equation}
thus implying that the ALP-originated photons that reach Earth from the SN are not necessarily emitted along the SN-Earth direction, but rather at an angle as schematically shown in Fig.~\ref{geometry}. Conversely, this implies that
on Earth we would see the photons as if they were coming from a direction somewhat off the location of the SN, i.e., the signal is effectively smeared out over a halo.

We must also consider that the length travelled by the ALP before decay plus the length travelled by the decay photon reaching Earth is longer than the distance between Earth and the SN, $d_{\rm SN}$~\cite{duffy} (cf. Fig.~\ref{geometry}). This leads to a time delay that, considering the distances involved, can be of the order of years even if the decay angles are not very large. The fact that massive ALPs have a velocity $< c$ increases the delay even further. Instead of a signal that lasts as long as the duration of the SN explosion and associated neutrino burst ($\sim 10$~s), as was the case for nearly massless ALPs in Refs.~\cite{Brockway,Toldra,Angelis, Simet, Serpico,ringwald} for SN 1987A, the signal from massive ALPs may be spread out over much longer time scales.


\subsection{Flux of massive ALPs} \label{sec_flux}

\indent

The subtleties of the core collapse of a progenitor star of mass around $\sim 18 M_{\odot}$ and the associated ALP production in its interior have been thoroughly analysed in Ref.~\cite{ringwald}. We adopt their results for the production rate of massless ALPs, which, for $g_{a\gamma\gamma} = 10^{-10} \, {\rm GeV}^{-1}$, can be fitted by
\begin{equation}
\frac{d\dot{N}_{a}}{dE_a} = a(t)E_a^{b(t)} \exp(-c(t)E_a) \label{prod}
\end{equation}
in overall units of $10^{50} \, {\rm MeV}^{-1}{\rm s}^{-1}$. The time-dependent coefficients, with adequate dimensions, are found to be $a(t) = 0.0054 - 0.001 \, t +5.77 \times 10^{-5} \, t^2$, $b(t) = 2.10 + 0.067 \, t - 0.004 \, t^2$ and $c(t) = 0.03 + 0.0003 \, t + 1.78 \times 10^{-5} \, t^2$, with $t$ in seconds and $E_a$ in ${\rm MeV}$. The total emission spectrum can be obtained by integrating this function over the time of the core collapse ($\sim 10$ s)~\cite{ringwald}, i.e.,
\begin{equation}
\frac{dN_{a}}{dE_a}=\int^{10\,{\rm s}}_{0}dt\,\frac{d\dot{N}_{a}}{dE_a}. \label{int_prod}
\end{equation}

\pagebreak

For our purposes, it is enough to consider the ALP production as instantaneous. 
We fit the integrated emission with the formula 
\begin{equation}
\label{approxflux}
\frac{dN_{a}}{dE_a}\bigg|_{\rm approx} = C \frac{E_a^2}{\exp(E_a/T)-1}\sigma_0(E_a),
\end{equation}
where $T$ is an effective temperature, $C$ contains information on the volume and density of scatterers and $\sigma_0(E_a)$ is the Primakoff production cross section for massless ALPs off non-relativistic targets given by
\begin{equation}
\sigma_0(E_a) = \frac{\alpha g_{a\gamma\gamma}^2}{8} \left[ \left( 1 + \frac{k_s^2}{4E_a^2} \right) \log\left( 1 +  \frac{ 4E_a^2}{k_s^2} \right) - 1 \right]. \label{cs_0}
\end{equation}
Here $k_s$ is an effective Debye screening scale that determines the range of the Coulomb potential created by the scattering centers in the plasma \footnote{If the scatterers in the plasma are non-relativistic, almost all the energy of the original photon is transmitted to the ALP, i.e.,  $\omega_\gamma\simeq E_a$.}. We find that Eq.~\eqref{approxflux} reproduces the integrated emission Eq.~\eqref{int_prod} to a good precision if we choose  $C = 2.54 \times 10^{77} \, {\rm MeV}^{-1}$, $T = 30.6$ MeV and $k_s = 16.8$ MeV.

\begin{figure}[t!]
\centering
\includegraphics[angle=0,width=0.57\textwidth]{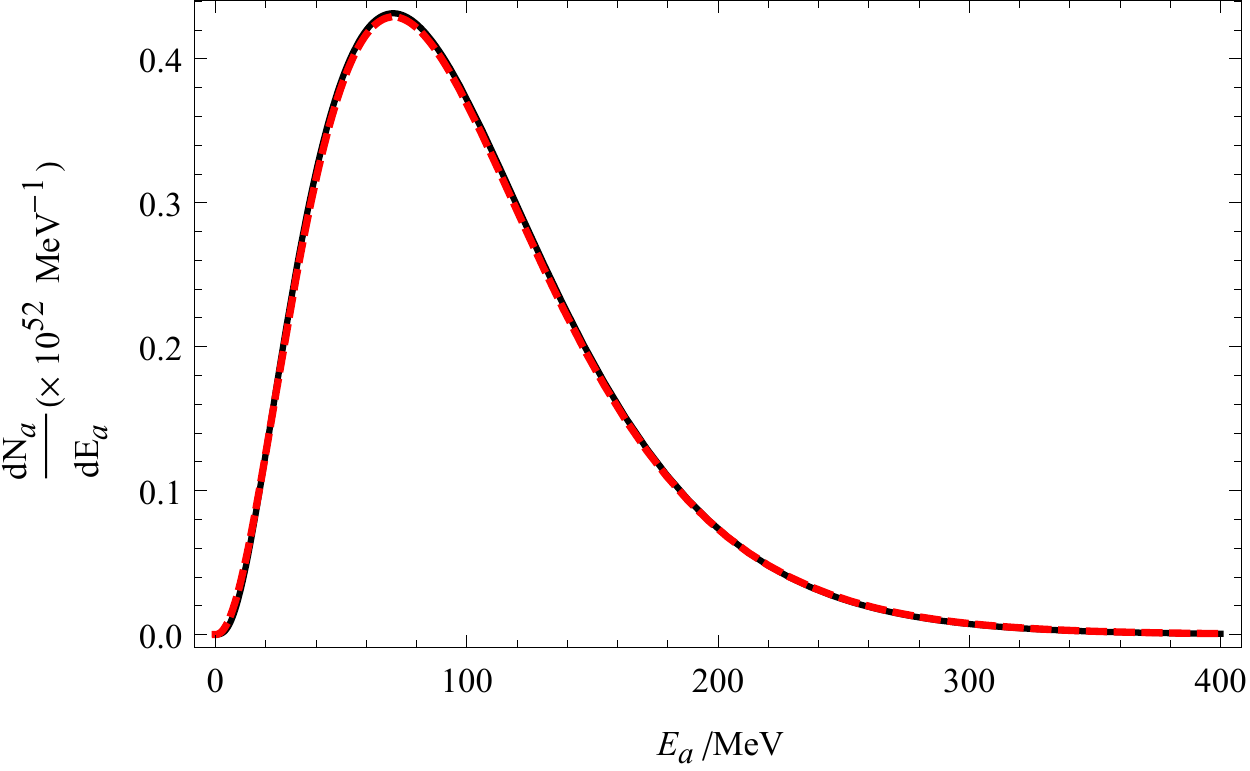}
\caption{Energy distributions from Eq.~\eqref{int_prod} (solid black) and Eq.~\eqref{approxflux}(dashed red).} \label{spec}
\end{figure}

Both results are compared in Fig.~\ref{spec}, which shows excellent agreement. This gives us confidence that our fit values are close to the relevant physical time and space averages.

The production rate for ALPs depends only on the properties of the plasma, i.e. the temperature and the Debye scale. This is true both for the massless, as well as for the massive case. In a real supernova both depend on time and location inside the core of the progenitor star. However, as we have just discussed, the full production in the massless case can be very well approximated by assuming a suitable effective temperature and Debye scale. We will therefore use this effective temperature and Debey scale to perform an extrapolation to the massive case. While this is certainly only an estimate, we believe it to be fairly accurate in light of the excellent fit for the massless case.

We can now calculate the ALP production by including the mass in the Primakoff cross section $\sigma_0$, which is found to be (cf. Ref.~\cite{cadamuro})
\begin{eqnarray}
\sigma(E_a) & = & \frac{\alpha g_{a\gamma\gamma}^2}{8} \Bigg\{ \left( 1 + \frac{k_s^2}{4E^2_a} - \frac{m_a^2}{2E_a^2} \right) \log\left[ \frac{ 2E_a^2(1+\beta) + k_s^2 - m_a^2 }{ 2E_a^2(1-\beta) + k_s^2 - m_a^2 } \right] - \beta  \nonumber \\
& - & \frac{m_a^4}{4k_s^2E^2_a} \log\left[ \frac{ m_a^4 + k_s^2 \left(2E_a^2(1+\beta) - m_a^2 \right) }{ m_a^4 + k_s^2 \left(2E_a^2(1-\beta) - m_a^2 \right) } \right]  \Bigg\}, \label{massive}
\end{eqnarray}
where $\beta = \sqrt{E_a^2 - m_a^2}/E_a$ is the ALP velocity in natural units.

The effect of non-vanishing ALP masses can now be encoded in an energy-averaged suppression factor, which we define as 
\begin{equation}
\mathcal{S}(m_{a})=\frac{\int\frac{d^{3}k}{(2\pi)^{3}}\frac{1}{\exp(\omega/T)-1}\sigma(\omega,m_a)}
{\int\frac{d^{3}k}{(2\pi)^{3}}\frac{1}{\exp(\omega/T)-1}\sigma_0(\omega)}, \label{sup}
\end{equation}
and show in Fig.~\ref{suppression} below. In the formula and figure we neglect the energy dependence for simplicity. However, in our numerical simulation this energy dependence is included. Note that ${\cal S}(m_a)$ is only appreciably different from one for $m_{a} \gtrsim \sqrt{k_s E_a}  \sim 20 \, {\rm MeV}$, cf. Eq.~\eqref{massive}, where we expect a suppression of the massive ALP production relative to the massless case. We can compute the ALP flux by rescaling the massless flux with this factor.

\begin{figure}[t!]
\centering
\includegraphics[angle=0,width=0.53\textwidth]{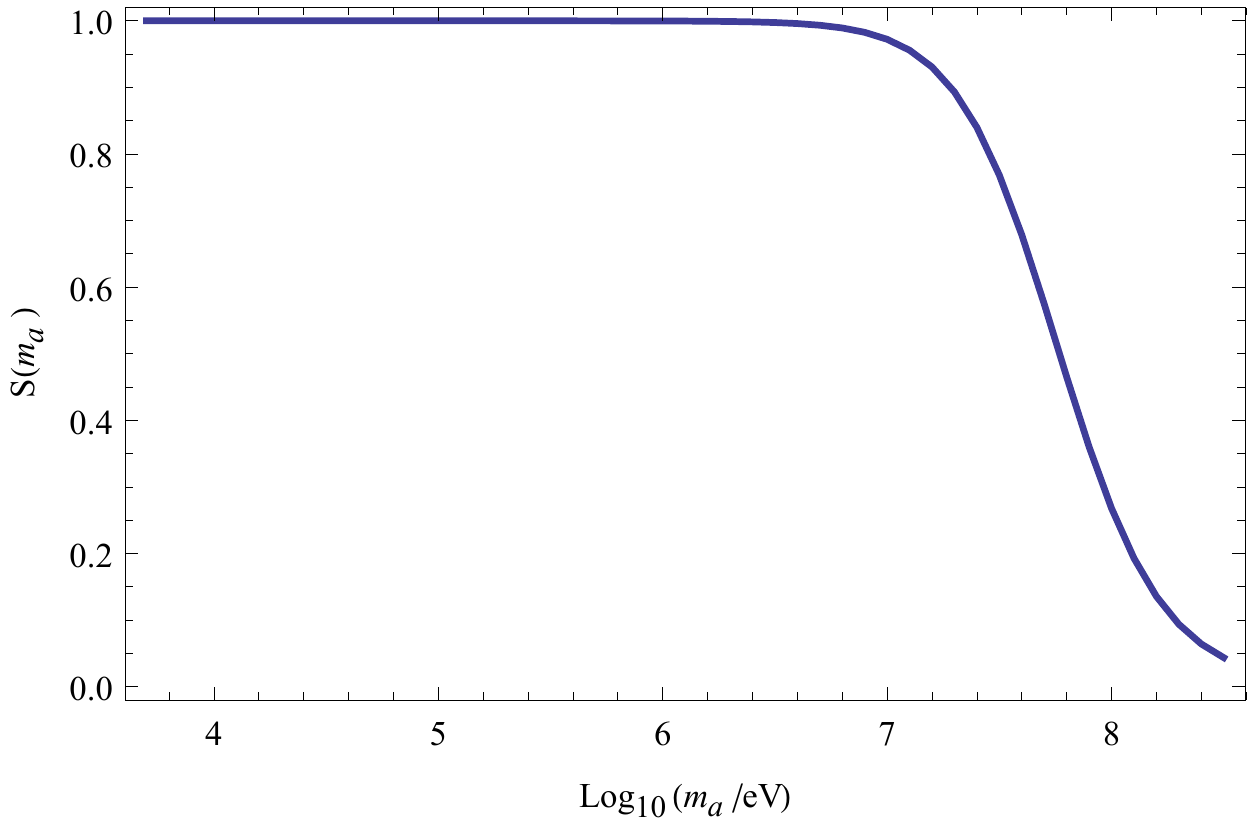}
\caption{Suppression factor, Eq.~\eqref{sup}, as a function of ALP mass. We assume the effective values $T = 30.6 \, {\rm MeV}$ and $k_s = 16.8\, {\rm MeV}$ for core temperature and Debye scale, respectively.} \label{suppression}
\end{figure}

\newpage


\subsection{Number of photons at the detector} \label{sec_num}

\indent

If all ALPs decay outside the SN, but before reaching Earth,
the effective fluence (i.e., the expected number of photons per unit detector area) $\mathcal{F}_\gamma$ of ALP-originated photons on Earth would be  
\begin{equation}
\label{flux}
\mathcal{F}_\gamma \big|_{\rm Earth} = \frac{2}{4\pi d_{\rm SN}^{2}} \displaystyle\int dE_a \left( \frac{dN_{a}}{dE_a} \right), 
\end{equation}
where $d_{\rm SN}$ is the SN-Earth distance and a factor of two is included to account for the two photons emitted per ALP decay. Here $dN_{a}/dE_a$, the ALP spectrum, is the result of integrating the production rate over the time of the core collapse ($\sim 10$ s)~\cite{ringwald}.

For nearly massless ALPs with instant decay we then have a naive fluence 
\begin{equation}
\label{events}
\mathcal{F}_{\gamma}^{\rm naive} = N_{\rm events}/{\rm cm^2} = \left( 3.3 \times 10^{6} \right) \frac{1}{\rm cm^2}\times \left( \frac{g_{a\gamma\gamma}}{10^{-10}\,{\rm GeV}^{-1}} \right)^2.
\end{equation}
However, as already indicated above, for massive ALPs with a finite decay length we have to take a variety of additional effects into account. Therefore we correct the fluence to 
\begin{equation}
\mathcal{F}_{\gamma}(m_a,g_{a\gamma\gamma}) = \mathcal{F}^{\rm naive}_{\gamma} \times \mathcal{P}_{\rm total}(m_a,g_{a\gamma\gamma}), \label{naive}
\end{equation}
where
\begin{equation} \label{prob_tot}
\mathcal{P}_{\rm total}(m_a,g_{a\gamma\gamma})  =  \mathcal{S}(m_a) {\mathcal{P}}_{\rm survival} {\mathcal{P}}_{\rm decay}{\mathcal{P}}_{\rm time}{\mathcal{P}}_{\rm acceptance}.
\end{equation}

Equations~\eqref{naive} and~\eqref{prob_tot} are only formal: the total probability and fluence cannot be factored in general, as photons with a given energy can originate from ALPs of different velocities depending on the decay angle. However, it shows the various effects that must be included in a realistic estimate of the sensitivity:
\begin{itemize}

\item{} $\mathcal{S}(m_a)$ is the mass-dependent factor defined in Eq.~\eqref{sup} which corrects Eq.~\eqref{approxflux} for massive ALPs. It is responsible for the suppression of the production for heavy ALPs ($m_a \gtrsim 20$ MeV).

\item{} ${\mathcal{P}}_{\rm survival}$ gives the fraction of ALPs decaying outside the region effectively occupied by the progenitor star, since the photons originating inside this region may be absorbed or at least scattered.

\item{} ${\mathcal{P}}_{\rm decay}$ takes into account that the ALP-originated photons from a decay at a distance greater than $d_{\rm SN}$ typically do not reach the detector. Photons emitted backwards with respect to the direction of the parent ALP could in principle reach the detector, even if the ALP decays after travelling beyond Earth, but their number is typically quite small so we neglect them in most of the discussion.

\item{} ${\mathcal{P}}_{\rm time}$ is the fraction of ALP-originated photons arriving within the measurement time of the detector. These are the only photons that can be effectively counted.

\item{} ${\mathcal{P}}_{\rm acceptance}$ accounts for the fact that some detectors may have a finite angular acceptance. Photons from ALP decays will arrive within a range of non-vanishing angles with respect to the SN (cf. Fig.~\ref{geometry}). Therefore, a detector with finite angular acceptance will only see a fraction ${\mathcal{P}}_{\rm acceptance}$ of all photons. Besides this, detectors usually have specific energy ranges where their sensitivity is optimal, e.g. for SN 1987A we work with photons in the $25-100$ MeV energy range at the detector \cite{Chupp,Woosley}; this cut is also described by ${\mathcal{P}}_{\rm acceptance}$.
\end{itemize}

In Section~\ref{simulation} we will numerically determine the effects of ${\mathcal{P}}_{\rm total}$, but it is worthwhile to briefly address the essential factors. To simplify the discussion we consider next a situation of an ALP with fixed energy and discuss the probability for the resulting photons to reach the detector. The probabilities for such a case will be denoted by $P$. To obtain the energy-averaged ${\mathcal{P}}_{\rm total}$ in Eq.~\eqref{events} the contributing factors have to be appropriately folded with the energy distribution (spectrum), which we chose to do numerically with a Monte Carlo simulation.

Since we already discussed $\mathcal{S}(m_a)$ earlier, let us comment on the second and third effects. The second factor is simply the survival probability for the ALP to leave the volume of the progenitor star before decay,
\begin{equation}
P_{\rm survival}=\exp\left[-\frac{R_{\star}}{\ell_{\rm ALP}(E_{a})}\right], \label{r_star}
\end{equation}
where $R_{\star}$, the effective radius, is much larger than the actual radius of the progenitor's core itself ($\sim$ 50 km for SN 1987A~\cite{ringwald}). Following Ref.~\cite{mohapatra} we shall take
\begin{equation}
R_{\star}\sim 3\times 10^{10}\,{\rm m}
\end{equation}
for SN 1987A. As can be seen from Eq.~\eqref{decaylength}, at large masses (above a few tens MeV) and couplings (above $\sim 10^{-10} \, {\rm GeV}^{-1}$), $\ell_{\rm ALP}$ is typically smaller than $R_{\star}$ so that, in this region, the sensitivity is strongly suppressed. For such large masses one may expect that the bound weakens due to the absorption of ALPs and this is indeed observed in our simulations (cf. Figs.~\ref{Exc_SN_223} and \ref{Exc_Bet}). On the upper-right corner one sees that the region where $\ell_{\rm ALP} \lesssim R_{\ast}$ is not excluded. The bound behaves as $g_{a\gamma\gamma} \sim 1/m_a^2$, which is compatible with Eq.~\eqref{r_star} (cf. Eq.~\eqref{decaylength}). We have conservatively assumed that any ALP decaying inside the effective radius of the progenitor star will be absorbed, thus not leaving an observable trace. Incidentally, the fact that decays within $R_{\star}$ are practically blocked inside the progenitor places an upper limit on the mass range available to our analysis.

The effect of the third and fourth factors are somewhat entangled. Let us consider the typical time delay of an ALP-originated photon. Compared to a massless particle travelling directly the distance between the SN and Earth, we observe a time delay
\begin{equation}
\Delta t = \frac{L_{1}}{\beta} + L_{2} - d_{\rm SN}. \label{time_delay}
\end{equation}

We are interested in the probability that the ALP decays before $d_{\rm SN}$ and is detected within a given time frame $\delta t$ (from data provided in Ref.~\cite{Chupp} we can extract limits for time intervals $\delta t \lesssim 220 \, {\rm s}$). For concreteness, we consider small masses ($m_a \lesssim$ MeV) and, in this region, we may estimate the fraction of events occurring within a measurement time $\delta t \sim 100$ s as follows. To incur a time delay $\Delta t \lesssim \delta t$ the ALP must decay before a distance $L_{\rm max} = \delta t\frac{\beta}{(1-\beta)} \sim \delta t \gamma^2 = \delta t \frac{E_a^2}{m^2_a}$. As long as $L_{\rm max} \lesssim \ell_{\rm ALP}$, the probability of measuring ALPs with time delays $\Delta t \lesssim \delta t$ is given by $P_{\rm decay} \times P_{\rm time} \approx L_{\rm max}/\ell_{\rm ALP}$, i.e.,
\begin{equation}
P_{\rm decay} \times P_{\rm time} \approx \delta t E_a \, g^2_{a\gamma\gamma}m_a^2
\end{equation}
which is the dominant effect limiting the sensitivity.

Having in mind that the ALP production cross section includes an extra factor of $g^2_{a\gamma\gamma}$, we see that, for a given detection time, the ALP-originated fluence $\mathcal{F}_{\gamma} = \mathcal{F}^{\rm naive}_{\gamma} \times \mathcal{P}_{\rm total}$ behaves as $\sim g^4_{a\gamma\gamma}m_a^2$, thus implying that the bound goes like
\begin{equation}
g_{a\gamma\gamma} \sim \frac{1}{\sqrt{m_{a}}}, \label{prob}
\end{equation}
which is exactly the behavior we observe in our numerical results (cf. Section~\ref{limits}).

Finally, since the detector for SN 1987A had a half-sky field of view~\cite{Chupp}, the angular acceptance has little impact on $P_{\rm acceptance}$. However, for SN 1987A we are considering the energy range $E_\gamma = [25,100]$ MeV, which imposes stronger constraints. As long as the ALP masses are not too large we have $E_\gamma \sim E_a/2$. Hence, the detector sensitivity corresponds to ALPs in the range $E_a = [50,200]$ MeV. Comparing with Fig.~\ref{spec} we may estimate $P_{\rm acceptance}$ as the normalized area under the curve within the aforementioned ALP energy range. By doing so we find that $P_{\rm acceptance} \sim 0.7$. As we shall see in Section~\ref{sec_bet}, for Betelgeuse $P_{\rm acceptance}$ will be reduced more significantly due to the energy range accessible to the Fermi-LAT detector that is slightly too high compared to the temperature of the supernova.


\subsection{Angular and time distributions} \label{sub_sec_C}

\indent

The angular and time distributions are closely related. Let us first note that, due to the assumed isotropy of the SN explosion, the angular and time-delay distributions will be the same at any point on a sphere with radius $d_{\rm SN}$ around the SN, cf. Fig.~\ref{geometry}. Therefore, to obtain the aforementioned distributions, it is enough to look at the distribution in angle and time with which the photons hit the surface of this sphere. In other words: if a photon originating from an ALP emitted in a certain direction hits this sphere, one can always find a rotation that puts Earth into the path of this particular photon. Hence, due to the isotropy assumption, emission of an ALP in this rotated direction has the same probability.

Let us consider an ALP emitted from the SN and decaying after covering a length $L_{1}$. One of the photons is emitted at an angle $\alpha$ relative to the direction of the parent ALP and then, after travelling a distance $L_2$, hits the sphere of radius $d_{\rm SN}$. From Fig.~\ref{geometry} we see that these quantities are related via $L_2^2 + 2\left( L_1\cos\alpha \right)L_2 + L_1^2 - d_{\rm SN}^2 = 0$, which can be solved for the distance travelled by the photon from the point at which the ALP decays,
\begin{equation}
L_{2} = -L_1\cos\alpha \pm \sqrt{d_{\rm SN}^2 - \sin^2\alpha L_1^2}. \label{L2}
\end{equation}
Moreover, from the law of sines one finds that the incidence (detection) angle of the ALP-originated photon with respect to the line of sight is given by\footnote{For a large number of photons with this detection angle the angular halo is $\Delta\phi \simeq 2\theta$.}
\begin{equation}
\sin\theta = \frac{L_1}{d_{\rm SN}} \sin\alpha. \label{sine}
\end{equation}

It is important to differentiate between two regions in space: $0 \leq L_1 \leq d_{\rm SN}$ and $L_1 > d_{\rm SN}$. For the former, corresponding to an ALP decaying between the SN and Earth, it is clear that, for $\alpha \leq \pi/2$, only the plus sign is physically meaningful. For an obtuse decay angle, due to the condition $L_1/d_{\rm SN} \leq 1$, the plus sign is again the only choice. Both situations imply an incident photon in the ``frontal" hemisphere of the detector, which is assumed to be aimed directly at the SN.

For the outer region, $L_1 > d_{\rm SN}$, a photon emitted with $\alpha \leq \pi/2$ will not be able to reach Earth, so only backward decays are relevant. 
Here, both signs may result in acceptable solutions provided that $\sin\alpha \leq d_{\rm SN}/L_1$. This condition is necessary to guarantee that the photon crosses the sphere with radius $d_{\rm SN}$ at least once. For angles satisfying this condition, the plus (minus) sign indicates the first (second) intersection of the photon with the sphere at $r = d_{\rm SN}$.

As already mentioned, the probability that a photon is emitted backwards is very small, since this can only happen if the parent ALP is not very boosted (either very heavy or low energetic) and, at the same time, travels beyond $r = d_{\rm SN}$. This is a highly unlikely scenario and most of the backward decays in the outer region do not reach Earth at all -- these photons are therefore essentially lost.

\bigskip

To get an idea of the size of the effects discussed above, let us evaluate the time delay for an ALP with $m_{a}=10 \,{\rm MeV}$ and $g_{a\gamma\gamma} = 10^{-10} \, {\rm GeV}^{-1}$. Taking $E_{a}=100\,{\rm MeV}$ as a representative value for its energy, the ALP emits a photon under an angle $\alpha \sim \gamma^{-1} \sim 6$ deg. Using $d_{\rm SN} = 51.4 \, {\rm kpc}$ for SN 1987A and assuming that the ALP decays after $L_1 \sim \ell_{\rm ALP} \sim 0.13 \, {\rm pc}$, Eqs.~\eqref{time_delay} and \eqref{L2} show that the overall time delay would be\footnote{Here the angular halo is $\Delta\phi \sim 10^{-5}$ deg (cf. Fig.~\ref{ang_win_sn}), which is small due to the very short decay length.} (cf. Fig.~\ref{time_hists})
\begin{equation}
\Delta t \sim 1.3 \times 10^3 \, {\rm s}. \label{ex_time_delay}
\end{equation} \indent
This example shows that the time delays may be significant, potentially spreading the signal over a period that is much longer than the duration of the SN explosion ($\sim 10 \, {\rm s}$). Repeating this exercise for points in the {\it allowed} region in parameter space shown in Fig.~\ref{ang_win_sn} we would get even larger effects. For $m_a = 1$ MeV and $g_{a\gamma\gamma} = 10^{-12} \, {\rm GeV}^{-1}$, we find that the time delay is $\Delta t \sim 3 \times 10^8$ s, whereas the angular halo is $\Delta\phi \sim 1$ deg (cf. Fig.~\ref{ang_dist}).

\bigskip
\bigskip

So far we have assumed a fixed emission angle of the photon with respect to the original ALP direction. Let us now justify this assumption. Since the ALP is a pseudo-scalar, in its rest frame, photon decay (emission) is equally likely in any direction, i.e., it is isotropic. Taking the Lorentz boost that brings the ALP from its rest frame into ours, where it travels with finite speed $\beta$, the originally isotropic angular distribution is distorted and is translated into an anisotropic one. To see this explicitly we consider the angular distribution for the separation angle between the two photons $ \psi = \alpha_1 + \alpha_2$. We find
\begin{equation}
\frac{dN_{\gamma_1 \, \gamma_2}}{d\psi} = \frac{1}{2\beta\gamma}\frac{\cos(\psi/2)}{\sin^{2}(\psi/2) }\frac{1}{\sqrt{\gamma^{2}\sin^{2}(\psi/2) - 1}}, \label{dist_ang}
\end{equation}
which is peaked at $\sin(\psi/2) = \gamma^{-1}$. In the $\beta \to 0$ limit,  the laboratory frame becomes also the ALP rest frame, with the photons emitted back to back, resulting in an increasingly peaked distribution around $\psi=\pi$.

The smallness of the typical decay angle for {\it both} emitted photons is the reason why only a very small fraction of the photons from ALP decays outside the sphere of radius $d_{\rm SN}$ around the SN can reach Earth. Backward emissions are therefore very unlikely already for relatively slow ALPs, justifying the previous comments.


\section{Simulation of the angular and time distributions} \label{simulation}


\subsection{Description of the simulation} \label{sec_desc_sim}

\indent

Approximate analytic results for the distribution in time have already been obtained\footnote{While we fully agree with the general approach from Ref.~\cite{duffy}, we were unable to reproduce their resulting limits. We think there are two reasons for that. 1) The approximation in their Eq. (2.10) requires $\frac{\Delta t}{d_{\rm SN}}\frac{1}{1-x\beta}\sim\frac{\Delta t}{d_{\rm SN}}\gamma^2 \ll 1$. For low masses and observation times several years later this does not seem to hold (in our case it does and we find a $g_{a\gamma\gamma} \sim m_a^{-1/2}$ behavior). 2) Emission with an effectively fixed temperature takes place only for a very small time frame $\sim 10$ s.} in Ref.~\cite{duffy}. We have instead used a full numerical simulation to account for the combined effect of the ALP production in the core of the SN, its motion out of the collapsing star and subsequent decay into two photons, as well as their path until arrival on Earth. Below we briefly describe  the simulation as well as the numerical results concerning the time-delay and angular distributions in the context of SN 1987A. We denote as ``valid events" the events that pass all the cuts and reach Earth, that is, these are the detected photons.

In order to estimate the size of these effects we first generate the ALP energy distribution (spectrum) for each $\{ m_a, g_{a\gamma\gamma} \}$-pair based on the massive Primakoff cross section, Eq.~\eqref{massive}. In our numerical simulation roughly $10^7$ ALPs are produced per $\{ m_a, g_{a\gamma\gamma} \}$-pair. For this value the results were stable. The $\{ m_a, g_{a\gamma\gamma} \}$ parameter space itself is scanned in steps of 0.1 (in log scale). This coarse graining produces slightly visible kinks in our limit which are however in line with the level of precision we are aiming for. Using the geometry displayed in Fig.~\ref{geometry}, we then sample for each ALP a distance $L_1$ (travelled by the ALP before it decays), which is exponentially distributed following $P_{\rm decay} = \exp\left[- L_1/\ell_{\rm ALP}(E_a)  \right]$. At this point we must impose the first physical cut by demanding that photons decaying inside the region $L_1 \leq R_{\ast}$ (cf. Eq.~\eqref{r_star}) are effectively absorbed and do not escape the SN, therefore not reaching the detector. This cut will only impact on relatively heavy and strongly coupled ALPs. For masses below a few tens of MeVs it has no significant effect.


After covering the distance $L_1$, the ALP decays in two photons. The decay is isotropic in the rest frame of the ALP, but not in the common Earth-SN frame. This can be taken into account by applying an appropriate boost to the ALP and its by-products, whereby we re-obtain the expected focusing of the decays in the forward direction, cf. Eq.~\eqref{dist_ang}. In the ALP's rest frame the ALP-originated photons have equal energies, $E_a/2$ each, but, due to the boost, in our frame their energies are distributed with some spread around this value. Since the detectors have in general a limited energy acceptance, we impose here our second important physical cut by limiting the valid events in the simulation to (boosted) photons with energies in the range $\left [ E_{-}, E_{+}  \right]$, with $E_{\pm}$ determined by the specific detector under consideration. In the case of the original measurements from SN 1987A, the optimal energy range was for gamma rays in the interval $25-100$ MeV~\cite{ringwald, Chupp}.


Both simulations, for the time delay and angular distributions, take the aforementioned aspects into account. We highlight again that, in our numerical simulations, the ALP production is taken as being instantaneous, i.e., all ALPs are produced at the same time in the core of the progenitor. We will return to this point in Section~\ref{limits}. Below we present a few representative examples, as well as discuss their most important physical features.


\subsection{Time distribution} \label{sec_time}

\indent

In Section~\ref{sub_sec_C} we discussed the path covered by the ALP and the ensuing photons and we showed that the combined trajectory leads to time delays often longer than the $\sim 10$ s duration of the neutrino burst associated with SN 1987A~\cite{Chupp}, cf. Eq.~\eqref{ex_time_delay}. As mentioned before, backwards photons reaching Earth are very rare, but these will be counted, despite of their relatively small contribution to the overall number of events.

\begin{figure}[!t]
\begin{minipage}[b]{0.41\linewidth}
\includegraphics[width=\linewidth]{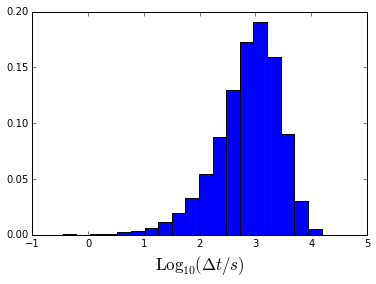}
\end{minipage} 
\begin{minipage}[b]{0.44\linewidth}
\includegraphics[width=\linewidth]{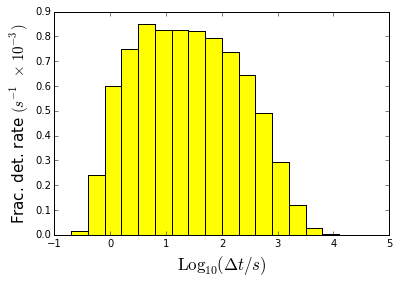}
\end{minipage}
\caption{For ALPs with $m_a = 10$ MeV and $g_{a\gamma\gamma} = 10^{-10} \, {\rm GeV}^{-1}$: fraction (probability) histogram depicting the distribution of time delays (left) and fractional detection rates (right).}
\label{time_hists}
\end{figure}

The time-delay simulation follows the steps indicated in Section~\ref{sec_desc_sim}: a number of ALPs is generated with the energy distribution from Fig.~\ref{spec} and travel a distance $L_1$, which is statistically determined by $\ell_{\rm ALP}$. In the sequence they decay into two photons that cover distances $L_2$ until detection (cf. Eq.~\eqref{L2}). The respective time delays -- two per ALP in general -- are then calculated according to Eq.~\eqref{time_delay}. An example of the distribution of time delays is shown on the left panel of Fig.~\ref{time_hists}, where time is in logarithmic scale for convenience.

Similarly, on the right panel of Fig.~\ref{time_hists} we show the {\it fractional} detection rate (in units of $10^{-3} \, {\rm s}^{-1}$), i.e., the fraction of ALP-originated gamma rays arriving at the detector per unit time\footnote{The height of each bin is given by the fraction of detections divided by the time length of that bin (note the logarithmic time scale).}. We see that the simple estimate leading to Eq.~\eqref{ex_time_delay} is sufficient to indicate the approximate time scale that marks the decline in the detection rate.

\subsection{Angular distribution} \label{sec_angular}

\indent

The incidence direction of the ALP-originated photons does not in general coincide with the line of sight between Earth and SN. It is therefore expected that the signal at the detector exhibits an angular spreading, i.e., a halo around the position of the SN, cf. Eq.~\eqref{sine}.

\begin{figure}[t!]
\centering
\includegraphics[angle=0,width=0.45\textwidth]{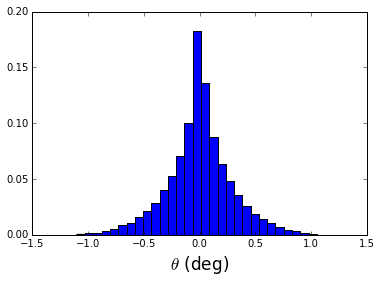}
\caption{Distribution for the incidence angle for photons originated from ALPs with $m_{a} = 1 \, {\rm MeV}$ and $g_{a\gamma\gamma} = 10^{-12} \, {\rm GeV}^{-1}$ emitted from SN 1987A. Here $\theta =\theta_{0}= 0$ corresponds to the direct line-of-sight between Earth and the SN.} \label{ang_dist}
\end{figure}

This halo could, in principle, cover a relatively large area of the sky\footnote{Here we only consider the (finite) angular region centered at the SN that contains a certain fraction, $f_{\rm ang}$, of the incident gamma rays. When discussing angular distributions, we do not consider particular time intervals, but rather show the effective angular windows after {\it all} ALP-originated photons have been detected.}. Considering the unconstrained angular distribution for a given $\{ m_a, g_{a\gamma\gamma} \}$-pair we obtain distributions similar to the one in Fig.~\ref{ang_dist}. This example shows that it is possible to reach sizeable maximal angular openings in the {\it non-excluded} region in parameter space -- in this case it is up to $\Delta\phi \sim 1^{\circ}$, which is about two times larger than the angular diameter of the Moon.

\begin{figure}[t!]
\centering
\includegraphics[angle=0,width=0.66\textwidth]{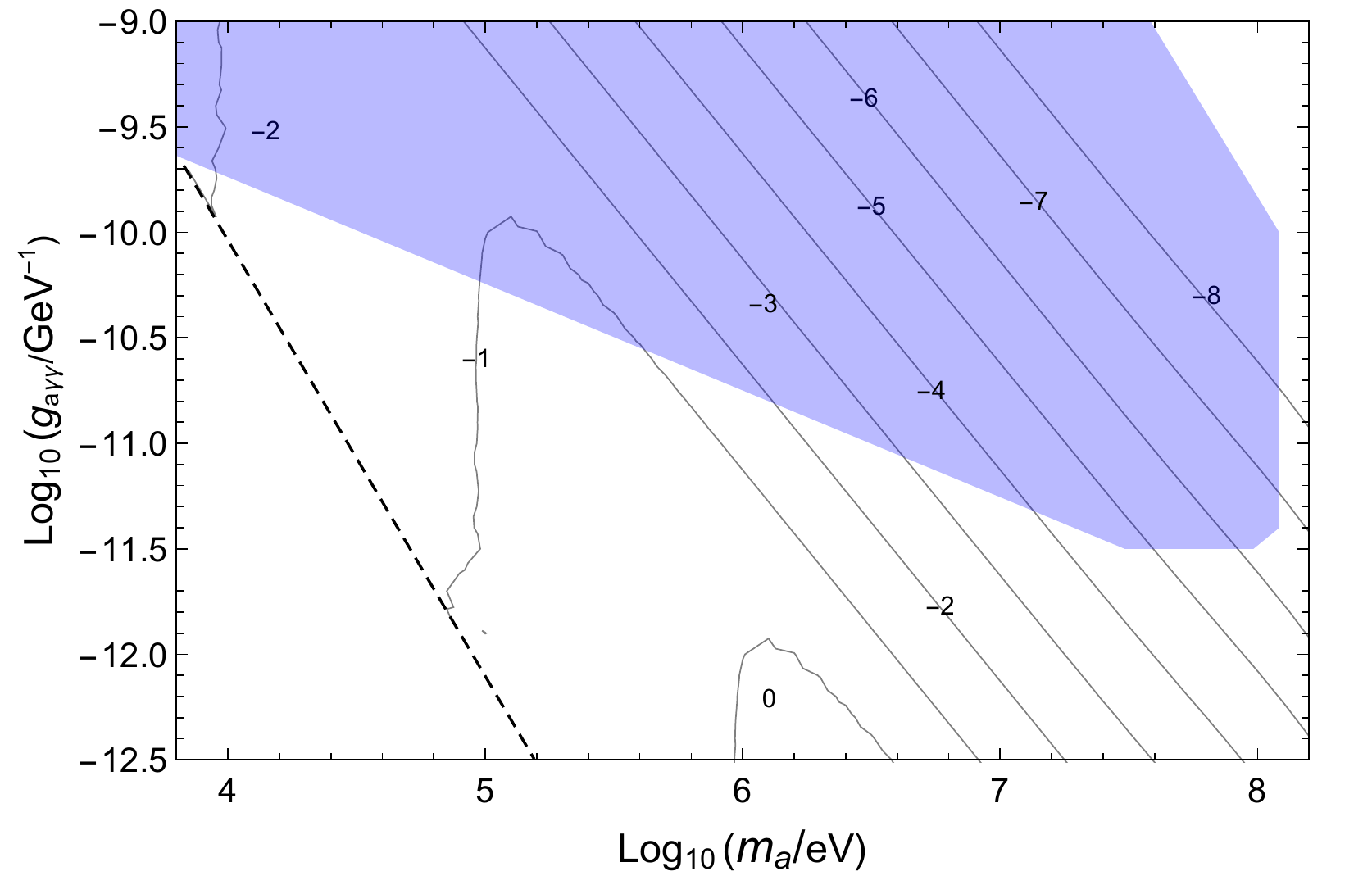}
\caption{{Angular size of the time-integrated photon glow from ALP decays} for SN 1987A. The contours correspond to the the angular interval containing $90\%$ of the ALP-originated photons arriving on Earth. We give values of  $\log_{10}(\Delta\phi/{\rm deg}$) close to the contours. The bound from Fig.~\ref{Exc_SN_223} for $\delta t \simeq 223$ s is shown in light blue. The dashed line indicates where our simulation gives essentially no events. We expect the vertical contours to continue beyond this point.} \label{ang_win_sn}
\end{figure}

The fraction of events within the angular acceptance is relevant to determine the sensitivity. Therefore, we are looking for the angular windows $\Delta\phi$ such that the interval $\left[ \theta_0  - \Delta\phi/2 , \theta_0  + \Delta\phi/2 \right]$ centered around $\theta_0$ contains the desired fraction $f_{\rm ang}$ of valid events. Here we have defined $\theta = \theta_0 = 0$ as the direction of line-of-sight to the SN. This is shown for SN 1987A in Fig.~\ref{ang_win_sn}, where we present the contours of constant angle (for convenience expressed in logarithmic scale as $\log_{10}\left( \Delta\phi/{\rm deg} \right)$) in the $g_{a\gamma\gamma} - m_a$ plane for $f_{\rm ang} = 90\%$.

Let us now discuss some features of Fig.~\ref{ang_win_sn}. The area on the lower-left corner (left of the dashed line) corresponds to the region of small masses and couplings. This region is not covered due to the extremely long decay lengths ($\ell_{\rm ALP} \gg d_{\rm SN}$), where very few ALP-originated photons are able to reach us. The lack of contour lines is not physically meaningful: it is an artifact due to the finite number of points in the numerical simulation.

It is important to notice that, as the decay length increases, the angular windows usually also get larger, but a considerable reduction in the overall number of events takes place. A similar effect happens for the photons emitted backwards: the angular windows are quite large, but the density of events is extremely low. For this reason, contrary to what we did in the time-delay simulations discussed previously, we refrain from including the signal from photons emitted backwards in the simulation for the angular distribution, as there are too few events.

The opposite behavior is observed in the central and upper-right areas, where masses are larger. There we have $\ell_{\rm ALP} \lesssim d_{\rm SN}$ and the majority of events is detected. Also important in this region is the fact that $\sin\theta \sim m_a^{-3}$ (cf. Eqs.~\eqref{decaylength} and \eqref{dec_ang}), so the angular windows are correspondingly tight. One must keep in mind, however, that in the large-mass region of the $m_a - g_{a\gamma\gamma}$ plane the decay length eventually shrinks below $R_{\ast}$, where the ALP-originated photons are effectively trapped inside the stellar material and do not leave.

The lines of constant detection angle follow $g_{a\gamma\gamma} \sim m_a^{-3/2}$, which is expected due to the combination $\sin\theta \sim \ell_{\rm ALP}\sin\alpha$, cf. Eq.~\eqref{sine}. It is noteworthy that, in the {\it allowed} regions -- those with ALP-originated fluence below the experimental upper limit -- it is possible to find relatively broad angular windows, reaching $\sim 10^{\circ}$ for large masses and small couplings.

Also interesting is the presence of vertical lines. To understand these, let us consider a fixed detection angle and a roughly fixed ALP energy. For heavy, not very boosted ALPs, the decay angles may vary within a relatively large range, so that many different paths will end up having the same detection angle -- even though they have a variety of travelled distances ($L_1 \sim \ell_{\rm ALP}$) and decay angles.

In a sense, the decay length compensates for the freedom in the decay angles. As one goes to smaller masses (i.e., the ALPs are more boosted), the decay angles are quickly more constrained, thus leaving less room for the decay length to compensate. In this low-mass region the detection angle is then dominated by the maximal decay angle $\sim \gamma^{-1}$, which is independent of the coupling constant (cf. Eq.~\eqref{dec_ang}), hence the saturation and vertical drop-off observed in Fig.~\ref{ang_win_sn}.

Moreover, in Fig.~\ref{ang_win_sn} we have superimposed the boundary of the excluded region where the ALP-originated fluence exceeds the experimental upper limit (cf. Fig.~\ref{Exc_SN_223}). We note that the maximal angular openings within the {\it excluded} region are $\sim 0.1^{\circ}$ in the $0.1 - 1$ MeV range.


\section{Limits from supernovae} \label{limits}


\subsection{SN 1987A} \label{sec_sn87}

\indent

The supernova from 1987, whose progenitor Sk -69 202 was a blue supergiant ($\sim~18 M_{\odot}$)~\cite{Woosley}, occurred at a distance of $d_{\rm SN} = 51.4$ kpc in the Large Magellanic Cloud. Its observation in visible light was preceded by a $\sim 10 \, {\rm s}$ long neutrino burst~\cite{Chupp}.

At the time of the event, the Gamma-Ray Spectrometer (GRS), which was sensitive in the $4.1 - 100$ MeV range with half-sky field of view, was mounted on the satellite-borne Solar Maximum Mission (SMM)~\cite{Chupp}. The analysis of the data showed that no excess of gamma-ray radiation reached the detector from the direction of SN 1987A during the neutrino burst. This non-observation was converted into an upper $3 \, \sigma$ limit on the fluence, namely, $\mathcal{F}_{\gamma}^{exp} < 0.6 \, \gamma \cdot  {\rm cm}^{-2}$ for photons in the energy band $25 - 100$ MeV.

As mentioned in Section~\ref{sec_desc_sim}, we assume that all ALPs are produced in the core of the progenitor at the same time. In practice, according to Ref.~\cite{ringwald}, this process happens in a time frame of $\sim 10$ s. In contrast to the nearly massless case studied in Refs.~\cite{Brockway,Toldra,Angelis, Simet, Serpico,ringwald,Osc}, for the photons from the decay of massive ALPs the time-delay distributions may be quite broad (cf. Sections~\ref{sub_sec_C} and~\ref{sec_time}).

It is therefore advantageous to use a longer time window after the first neutrino recorded. To do so we look at the full time window of Ref.~\cite{Chupp}, $\delta t \simeq 223$ s, and consider the $3 \, \sigma$ statistical fluctuation on the fluence in this period. Since no excess number of events are recorded compared to the control region we use $N = 1393$ events and the estimate $\sigma = \sqrt{N}$. The upper bound on the fluence for the extended observation time is $\mathcal{F}_{\gamma}^{ {\rm exp} } (223 \, {\rm s}) \leq 3 \times \sigma / A_{\rm eff} = 1.78 \, \gamma \cdot {\rm cm}^{-2}$, with the effective area $A_{\rm eff} = 63 \, {\rm cm}^2$~\cite{Chupp}. This effective area takes into account the full field of view of the detector. In the region in parameter space under considerationr the angular distribution is still very narrow -- as in the case of photons from ALP conversions in magnetic fields -- and we simply follow the same procedure as in Ref.~\cite{ringwald,Chupp}.

\begin{figure}[t!]
\centering
\includegraphics[angle=0,width=0.66\textwidth]{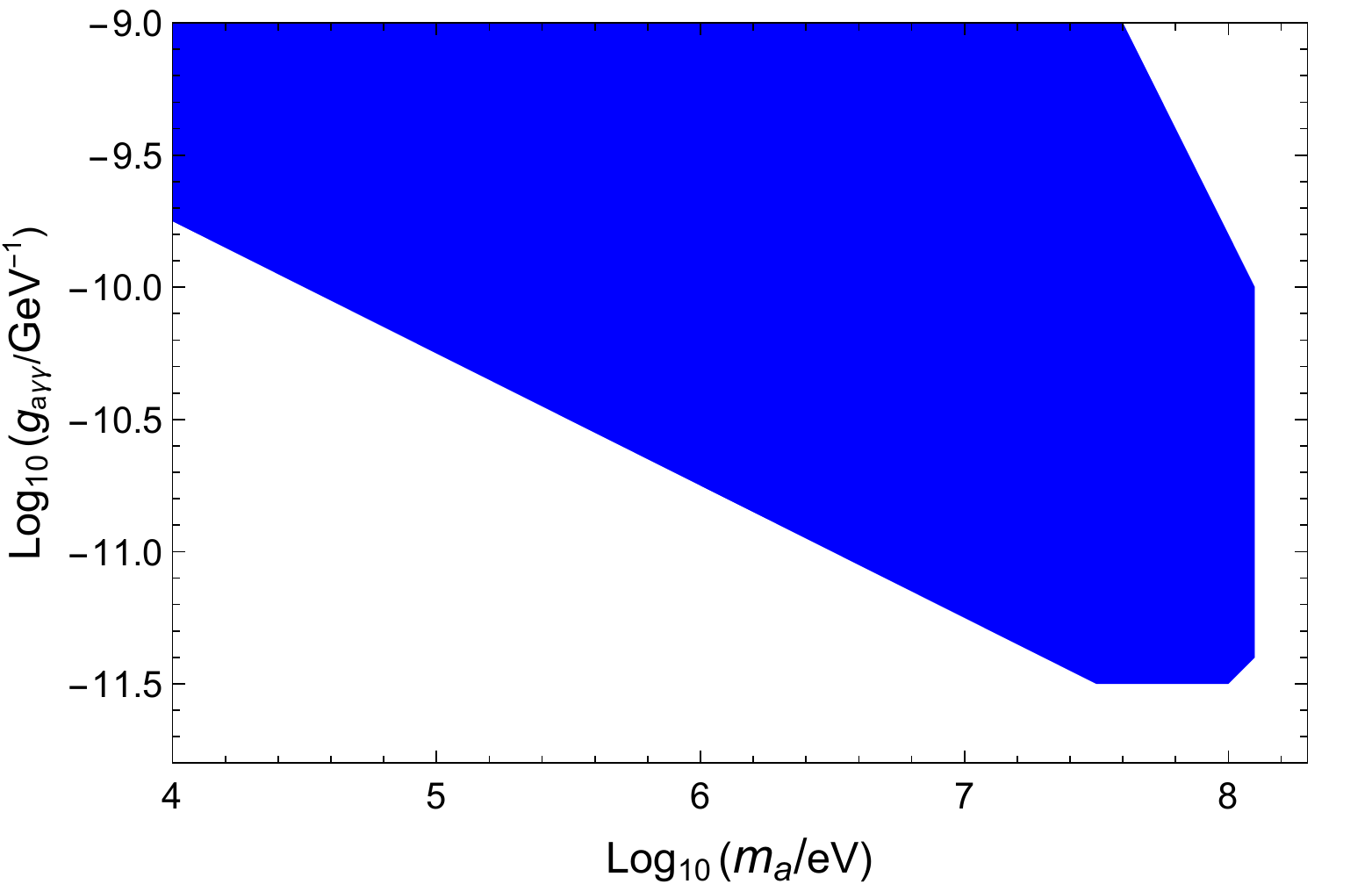}
\caption{Bound based on the fluence for SN 1987A with $\mathcal{F}_{\gamma}^{ {\rm exp} } (223 \, {\rm s}) < 1.78 \, \gamma \cdot {\rm cm}^{-2}$. The excluded region is displayed in blue.} \label{Exc_SN_223}
\end{figure}

With this upper limit on the fluence we are able to derive the bound presented in Fig.~\ref{Exc_SN_223} that is also included in the overview Fig.~\ref{exc_plot}. As mentioned in Section~\ref{sec_num}, the shape of the bound in the low-mass region behaves as $g_{a\gamma\gamma} \sim m_a^{-1/2}$. The non-excluded region increases as the mass decreases due to the decay length: as the masses get smaller, the ALPs are able to survive statistically longer, until the point where they decay predominantly behind Earth, so that, being extremely boosted, very few photons reach us on average, thus suppressing the bound accordingly.

\newpage

The above behavior is sustained until masses of order $m_a \sim 20$ MeV and then a turn-up takes place, as anticipated in the discussion of the suppression factor, cf. Fig.~\ref{suppression}. This is attributed to the size of the physical parameters entering the (massive) Primakoff cross section, Eq.~\eqref{massive}, namely, the effective temperature and the Debye screening scale, which are both $\mathcal{O}(10 \, {\rm MeV})$. This is the point where the production rate ``feels" that the ALPs in the final state are actually heavy, thus consuming a portion of the energy available to convert it into rest mass for the ALPs. This reduces the number of ALPs produced and the fluence is correspondingly suppressed, thus causing the bound to recede.

Another interesting feature is the impact of the effective radius, which is visible on the upper-right region of Fig.~\ref{Exc_SN_223}. This region is characterized by very small decay lengths $\ell_{\rm ALP} \lesssim R_{\ast}$. In this region the ALP-originated photons are absorbed and cannot be detected on Earth, thus explaining the {\it allowed} region on the upper-right area (cf. Section~\ref{sec_num}).

The energy range of the photons used to obtain the bound in Fig.~\ref{Exc_SN_223} is $25 - 100$~MeV. This is more or less the optimal range. Since each ALP decays in two photons, the energy of each photon is distributed around $E_a/2$ and, given that the ALP spectrum achieves a maximum around $E_a \sim 80$ MeV, most of the ALP-originated gamma rays will be produced with $E_{\gamma} \sim 40$ MeV. In this sense, the optimal energy range for gamma-ray detection should include this value -- and possibly even lower ones -- in order to cover the range where ALP production is largest.

Let us briefly comment on the uncertainties involved in our limit. There are two uncertainties related to the production process of massive ALPs. One is our simplified modelling of the plasma with an effective temperature and Debye scale. As we have argued in Section~\ref{sec_flux},  we think that this is a relatively small effect. Moreover, there are uncertainties in the modelling of the supernova itself. These have been discussed in Ref.~\cite{ringwald} and is assumed to be also reasonably small. In our figures we include only the statistical $3\sigma$ uncertainty, but the systematic uncertainties should be kept in mind.


\subsection{Betelgeuse} \label{sec_bet}

\indent

Betelgeuse is a red supergiant with similar mass as the progenitor from SN 1987A, $\sim 18 \, M_{\odot}$, located in the Orion constellation around 200 pc (650 ly) from Earth. It is expected to explode in a SN event in the next few hundred thousand years~\cite{bet2}. Due to its proximity, it is one of the brightest objects in the night sky and, should it transition into a SN, its associated ALP-originated gamma-ray flux would be much more intense than the one from SN 1987A. Besides this, the gamma-ray instruments have improved in the last decades, so we expect that the overall sensitivity will be significantly better, thus allowing us to set stronger bounds on the ALP parameter space.

Currently, one of the best detection possibilities would be the Fermi-LAT, whose point-source sensitivity after an observation time of one year ($\sim 3 \times 10^7$ s) is $3 \times 10^{-9} \, \gamma \cdot {\rm cm^{-2}} \cdot {\rm s^{-1}}$ for incident photons with energies $E_{\gamma} > 100 \, {\rm MeV}$~ \cite{Fermi}.

\begin{figure}[t!]
\centering
\includegraphics[angle=0,width=0.66\textwidth]{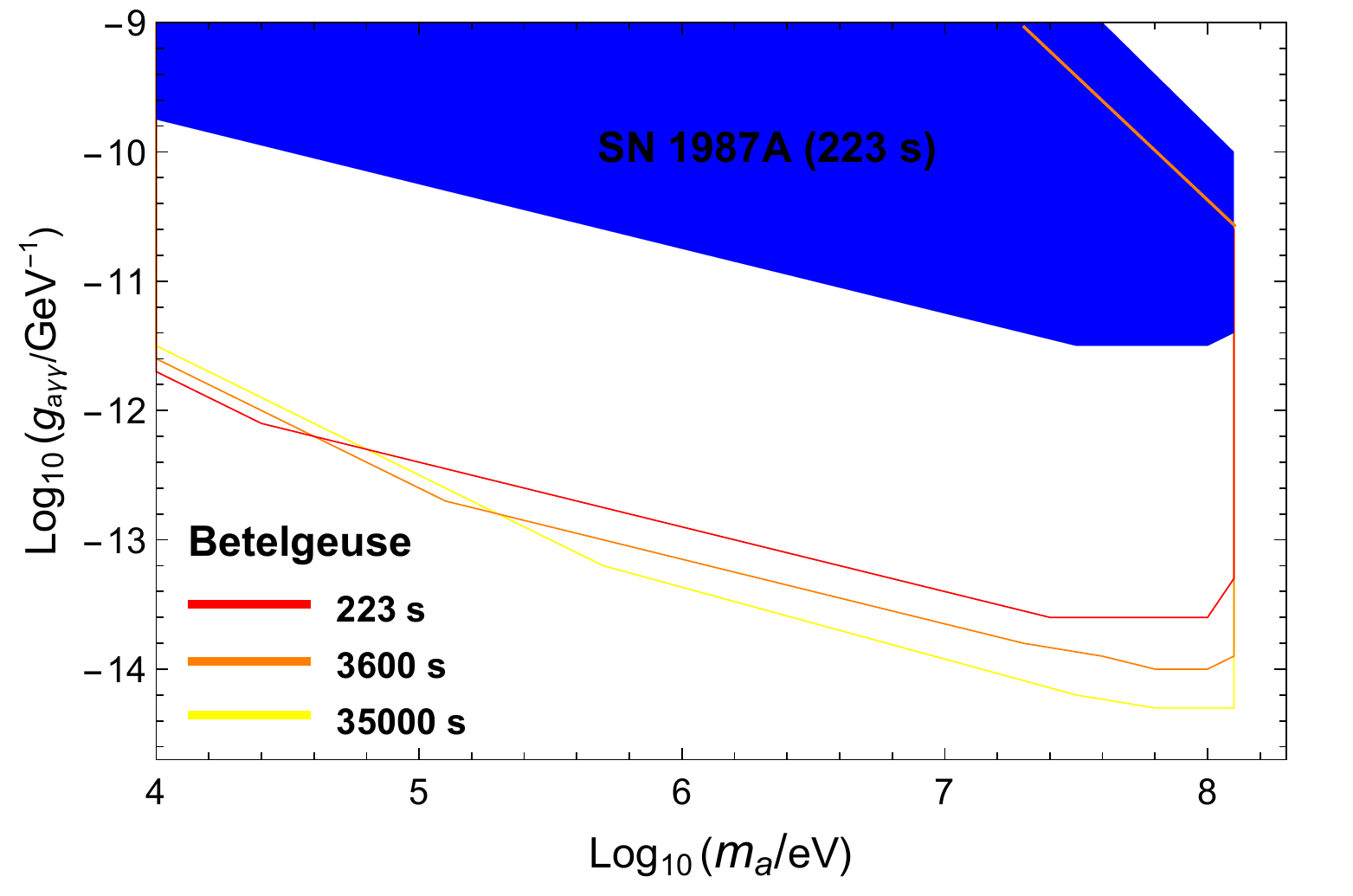}
\caption{Bounds based on the 223 s, 3600 s and 35000 s observation times for Betelgeuse using the effective area of the Fermi-LAT. The bound from SN 1987A is shown for comparison.} \label{Exc_Bet}
\end{figure}

The point-source sensitivity of the Fermi-LAT arises from a background flux given in Ref.~\cite{flux} together with the angular resolution~\cite{resolution} and the observation time of the point source.
For our purposes we consider a background flux of~\cite{flux}
\begin{equation}
{\rm background\,\,flux} = 1.5\times 10^{-5}\frac{1}{{\rm cm}^2\,{\rm s}\,{\rm sr}}.
\end{equation}
together with a conservative angular resolution of 5 degrees in all directions. Using this we obtain a detectable photon fluence over the background (assumed to be stable) at the 3~$\sigma$ level.
Taking the effective area of $9500\,{\rm cm}^2$ into account we find $1.9 \times 10^{-3}\gamma\/{\rm cm}^2$ for an observation time of $35000\,{\rm s}$, $6.2 \times 10^{-4}\gamma/{\rm cm}^2$ for $3600\,{\rm s}$ and $3.0 \times 10^{-4}\gamma/{\rm cm}^2$ for $223\,{\rm s}$. For the smallest time frame of $223\,{\rm s}$ there is effectively no background and we have taken the detectable number of photons to be 3. The corresponding bounds are displayed in Fig.~\ref{Exc_Bet}.

%
%
%
%
%

The bounds shown in Fig.~\ref{Exc_Bet} behave as expected in the $\sim 1 - 25$ MeV-mass region. However, for masses below $\sim 1$ MeV, the $g_{a\gamma\gamma} \sim m_a^{-1/2}$ behavior changes to $g_{a\gamma\gamma} \sim m_a^{-1}$. This is due to two intertwined factors: the relatively shorter distance to Betelgeuse and the observation times.

To see how these factors give rise to the observed behavior we have to remember that $\ell_{\rm ALP} \gg d_{\rm SN}$ for such small masses and couplings. This means that only ALPs decaying with $L_1 \lesssim d_{\rm SN}$ result in successful detections. Therefore, the detection probability is dominated by the probability of ALP decay essentially at $\sim d_{\rm SN}$, i.e., $P_{\rm decay} = 1 - \exp\left( -d_{\rm SN}/\ell_{\rm ALP} \right) \approx d_{\rm SN}/\ell_{\rm ALP}$. With Eq.~\eqref{decaylength} and the extra $g^2_{a\gamma\gamma}$ from ALP production this gives $\mathcal{F}_\gamma \sim g^4_{a\gamma\gamma}m_a^4$, which leads to 
\begin{equation}
g_{a\gamma\gamma} \sim 1/m_a. \label{g_bet}
\end{equation}

From Fig.~\ref{Exc_Bet} we see that the change in behavior takes place at different points for different observation times. It is therefore important to estimate where these changes occur. Firstly, we notice that we are limited to a finite observation time $\delta t$. Hence, the ALP-originated photons that are effectively counted at the detector need to arrive within this time period, i.e., $\Delta t \leq \delta t$ and the maximal time delay allowed is $\delta t$. Since $m_a \ll E_a$ we have very boosted ALPs and practically collinear photons, cf. Eq.~\eqref{dec_ang}. This means that $L_2 \approx d_{\rm SN} - L_1$ and we may write Eq.~\eqref{time_delay} as $\Delta t \approx L_1 \frac{1-\beta}{\beta} \approx L_1 \frac{m_a^2}{2E_a^2}$ or, more conveniently, 
\begin{eqnarray}
\Delta t & \approx & 5.2 \times 10^{-6} \, {\rm s} \, \left( \frac{m_a}{\rm eV} \right)^2 \left( \frac{L_1}{\rm kpc} \right) \left(  \frac{100 \, {\rm MeV}}{E_a} \right)^2. \label{dt}
\end{eqnarray}

We limit ourselves to time delays of at most the observation time, i.e., $\Delta t = \delta t$, which happen for $L_1 = d_{\rm SN}$. Plugging this into Eq.~\eqref{dt} we find that the mass $\tilde{m}$ which marks the transition is approximately given by $\tilde{m} \approx 4.4 \times 10^2 \, {\rm eV} \, \left(  \frac{E_a}{100 \, {\rm MeV}} \right) \sqrt{ \left(\frac{\rm kpc}{d_{\rm SN}} \right) \left( \frac{\delta t}{\rm s}  \right) }$. The fact that $\tilde{m}$ increases with $\delta t$ is reasonable, since heavier ALPs move slower, so longer observation times are sensitive to larger masses. One must note however that here $E_a$ must be such that the photon energies satisfy the constraints of the detector (for $E_a \gg m_a$ we have $E_a \sim 2E_\gamma$). In fact, this expression roughly matches the transition points for Betelgeuse with the Fermi-LAT detector shown in Fig.~\ref{Exc_Bet}. For SN~1987A, however, $\tilde{m} < 10$ keV and the low-mass behavior is not visible\footnote{Comparing with Ref.~\cite{duffy} with the observation 22 years after the SN event, we find that~$\tilde{m} \sim 1$ MeV.} in Figs.~\ref{Exc_SN_223} and~\ref{Exc_Bet}.

\begin{figure}[t!]
\centering
\includegraphics[angle=0,width=0.66\textwidth]{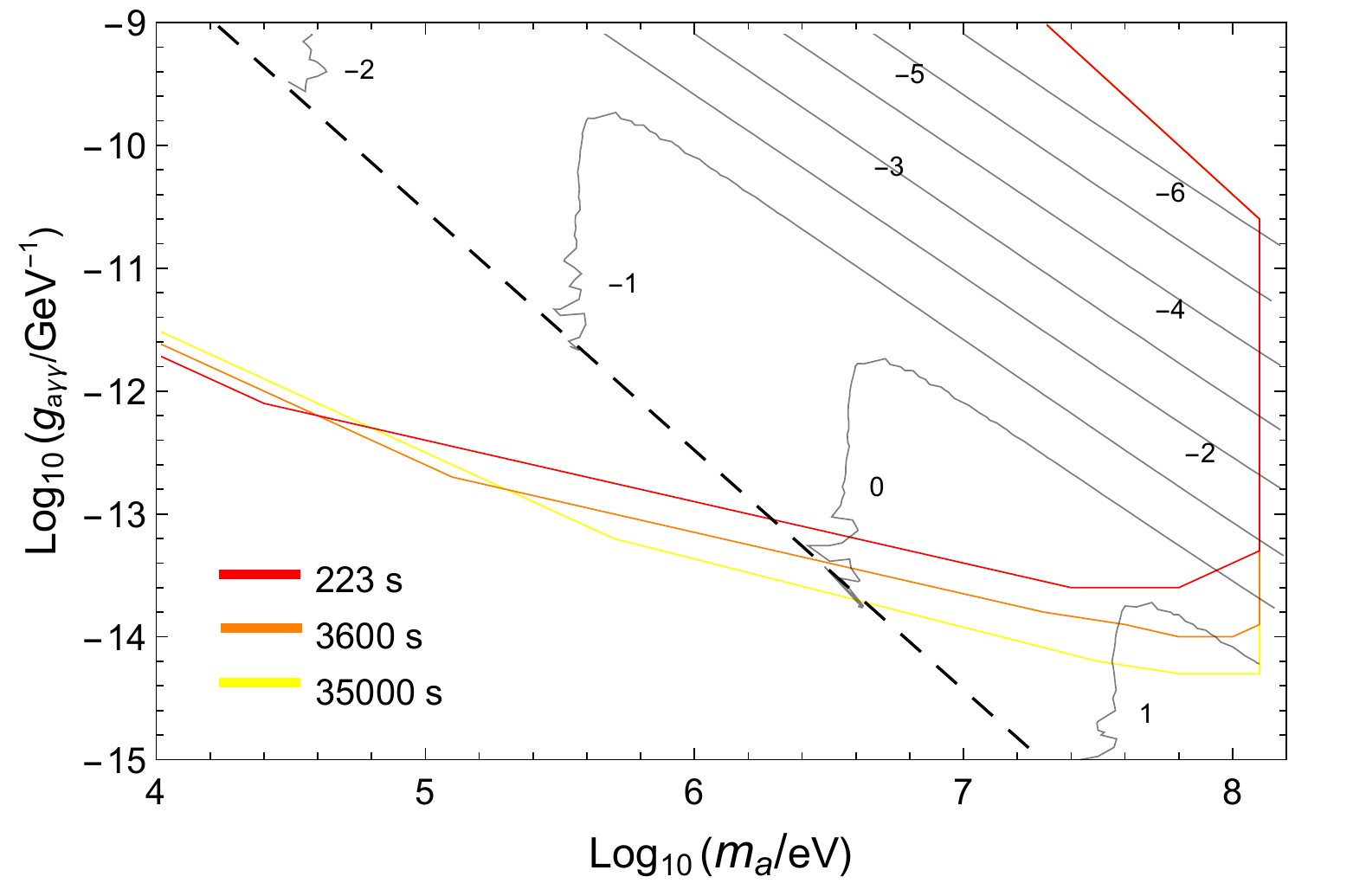}
\caption{Angular windows for Betelgeuse. The contours correspond to the angular interval containing $90\%$ of the ALP-originated photons arriving on Earth. The bounds from Fig.~\ref{Exc_Bet} for the different observation times are shown in gray scale.} \label{ang_win_bet}
\end{figure}

One may wonder why for very small masses an increased observation time does not lead to an improvement in the limit (cf. Fig.~\ref{Exc_Bet}), but instead actually to a slight weakening. The reason for this is rather simple. In this region the time delay is actually quite small and all photons that will ever arrive already do so in a time smaller than the smallest chosen observation time. Due to the presence of a non-vanishing background for longer observation times we have more background events without any gain in signal events, hence the limit is weaker.

Furthermore, the anticipated effect of $R_{\ast}$ is visible in the upper-right corner, where it limits the excluded region in a similar way as for SN 1987A. One must also note that the effective radius for Betelgeuse is $\sim 20$ times larger than for SN 1987A. This results in the corresponding worsening of the bounds in that region.

The projected improvement for the sensitivities from Betelgeuse is due basically to two factors: the larger effective area of the detector -- $9500 \, {\rm cm}^2$ for Fermi-LAT~\cite{Fermi} compared to $63 \, {\rm cm}^2$ from the SMM~\cite{Chupp} -- which leads to a lower upper limit on the fluence, and the shorter distance to Earth ($d_{\rm SN} = 0.2 \, {\rm kpc}$ for Betelgeuse, in comparison with $d_{\rm SN} = 51.4 \, {\rm kpc}$ for SN 1987A). This also causes the displacement of the constant-angle contours in Fig.~\ref{ang_win_bet}. For a given $\{m_a, g_{a\gamma\gamma} \}$-pair, the angular window for Betelgeuse is expected to be $\sim 200$ larger than for SN 1987A, cf. Eq.~\eqref{sine}.

Furthermore, the angular acceptance of the Fermi-LAT detector does not strongly constrain $P_{\rm acceptance}$. However, as mentioned in Section~\ref{sec_num}, this factor also takes the energy range of the detector into account: for Fermi-LAT we have $E_\gamma > 100$ MeV. Keeping in mind that $E_\gamma \sim E_a/2$, from Fig.~\ref{spec} we see that $E_a \gtrsim 200$ MeV is far from the peak of the ALP production, thus causing the sensitivity to drop by a factor of $P_{\rm acceptance} \approx 0.06$. Future experiments like e-ASTROGAM~\cite{eASTRO}, ComPair~\cite{compair}, or PANGU~\cite{pangu} will hopefully be able to improve on this aspect. Yet, even with this reduction, the ALP-originated gamma-ray flux from Betelgeuse is significantly larger due to the closer distance.

As a final remark, we would like to note that a possible ALP burst from Betelgeuse going supernova should not be dangerous to us on Earth. Recent analyses indicate that Betelgeuse would release $\Delta E \sim 10^{53}$~erg of energy -- similar to SN~1987A --, but the X- and gamma-ray emissions would not be large enough to penetrate Earth's atmosphere~\cite{Dolan}. On the other hand, the ALP-related energy release is $\Delta E_{\rm ALP} \sim 10^{50} \, {\rm erg} \left( \frac{g_{a\gamma\gamma}}{10^{-10} \, {\rm GeV}^{-1}} \right)^2$, which is much smaller that $\Delta E$. Hence, we do not expect the ensuing ALP-originated gamma-ray burst to pose any harm to us on Earth.


\section{Concluding remarks} \label{conclusion}

\indent

In this paper we have derived new limits on massive ALPs purely coupled to two photons from the supernova explosion SN 1987A. The process we use for our limits is the Primakoff production inside the core of the collapsing progenitor and subsequent decay into two photons. In the 10 keV - 100 MeV mass range these limits improve upon existing laboratory and astrophysical limits (see Fig.~\ref{exc_plot}).

Our limits overlap with the cosmological limits discussed in Refs.~\cite{cadamuro,Millea}, which are based on the effects of the decay of early universe relic ALPs on several CMB and BBN observables. While cosmological limits from thermal production of ALPs in the early Universe are potentially stronger, they are model dependent.

The grey region of Fig.~\ref{exc_plot} is excluded assuming that ALPs were in thermal equilibrium with the rest of primordial plasma and that the expansion of the universe afterwards was dominated by the relativistic degrees of freedom of the SM. These assumptions set the relic abundance of ALPs and require a sufficiently large maximum temperature of the Universe,  
\begin{equation}
T_{\rm RH} > T_{\rm fo} \sim 123 \,{\rm GeV} \frac{\sqrt{g_*}}{g_q}\left(\frac{10^{-9}\rm GeV}{g_{a\gamma\gamma}}\right)^2,
\end{equation}
where $g_*$ and $g_q$ are the energy and electric charge effective number of relativistic species, respectively (see Ref.~\cite{cadamuro}). Of course, achieving such a large temperature depends on the cosmological model considered. 
Many models feature lower reheating temperatures, which would not be large enough to produce the thermal abundance assumed in the constraints.

A very conservative lower limit is set by standard BBN, which requires $T_{\rm RH}\gtrsim 20$ MeV, implying that only for $g_{a\gamma\gamma} > 7 \times 10^{-8} \, {\rm GeV}^{-1}$ the bounds are independent of the cosmological model. The constraints discussed in this paper, even if superficially weaker, do not suffer from this model dependence and imply a more robust exclusion.

As already discussed in Ref.~\cite{duffy}, a large fraction of the ALP-originated photons arrives significantly delayed compared to the neutrinos from the $\sim 10$ s-long burst. Indeed, depending on the parameter values, we have found that the signal can be spread out over years. Although this dilutes the signal, it also provides opportunities as photons may be observed today or even in the near future with better instruments than were available in 1987.

\begin{figure}[t!]
\centering
\includegraphics[angle=0,width=0.66\textwidth]{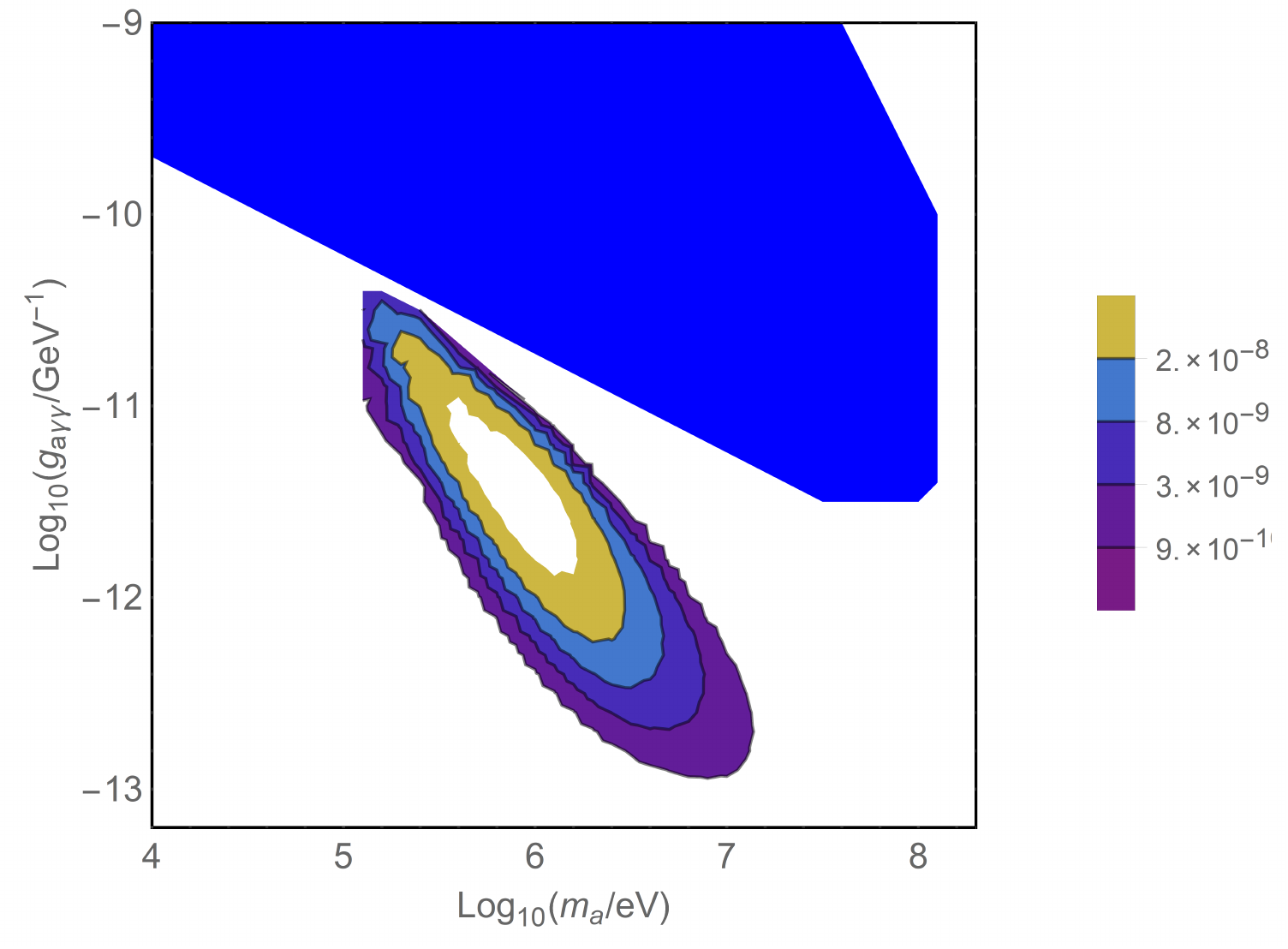}
\caption{The contour plot shows the average flux (in units of $\gamma \cdot {\rm cm^{-2}} \cdot {\rm s^{-1}}$ ) of ALP-originated photons from SN~1987A between 30 and 40 years after its explosion with detection energy $E_{\gamma}\geq 5\,{\rm MeV}$. We see that for a detector with the same point-source sensitivity as Fermi-LAT ($3 \times 10^{-9} \, \gamma \cdot {\rm cm^{-2}} \cdot {\rm s^{-1}}$), but with a significantly lower energy threshold, a sizeable area of parameter space could be accessed. For comparison we show in blue the limit obtained from the GRS, cf. Fig.~\ref{Exc_SN_223}.}  \label{fluxtoday}
\end{figure}

In fact, we have also looked for a possible ALP-originated signal from SN~1987A between today (i.e., 30 years after) and ten years from now. Unfortunately, for Fermi-LAT with its energy cut of $E_\gamma > 100 \, {\rm MeV}$ we have found that no signal is detectable. This is mainly due to the high energy required for photons to be accepted in that detector. This requirement reduces the signal twofold, as the SN-ALP spectrum is strongly suppressed at energies $\gtrsim 200\,{\rm MeV}$ and the highly energetic ALPs are typically less delayed and arrive earlier (i.e. before today).

If, however, we lower the energy cut to, say, $E_\gamma > 5 \, {\rm MeV}$, it is possible that some ALP-originated photons reach us in the next ten years. In Fig.~\ref{fluxtoday} we show the average flux of ALPs between 30 and 40 years after the SN~1987A event. With such an improved gamma-ray detector a considerable amount of interesting parameter space could be probed.

Notably, along with the time delay, the signal will also be spread in angles away from the line of sight. While this effect is not very large for SN 1987A, it can become important for future, closer, supernovae which may be observed with gamma-ray instruments with better directional readiness and angular resolution.

As an example we discuss the possibility of Betelgeuse going supernova in the future, leading to a significantly improved sensitivity. We view this as good motivation to investigate potential improvements also in other limits based upon SN 1987A that could be improved by a future SN observation and also consider the experimental ``readiness'' to maximally exploit such an opportunity.


\begin{acknowledgements}
We would like to thank M. Giannotti, G. Raffelt, C. Weniger and the anonymous referees for very helpful comments and suggestions. J.J. gratefully acknowledges support by the DFG TR33 ``The Dark Universe'' as well as the European Union Horizon 2020 research and innovation under the Marie Sklodowska-Curie grant agreement Numbers 674896 and 690575. P.C.M. would like to thank the Brazilian National Council for Scientific and Technological Development (CNPq) and the German Service for Academic Exchange (DAAD) for the invaluable financial support and the Institut f\"ur theoretische Physik (U. Heidelberg) for the hospitality. J.R. is supported by the Ramon y Cajal Fellowship 2012-10597 and FPA2015-65745-P (MINECO/FEDER).
\end{acknowledgements}


\end{document}